\documentclass{ws-mpla}

\begin{document}

\renewcommand{\draftnote}{} 
\renewcommand{\trimmarks}{} 

\markboth{E. A. Matute} {Restoration of Parity Symmetry through
Presymmetry}

\catchline{}{}{}{}{}

\title{RESTORATION OF PARITY SYMMETRY THROUGH PRESYMMETRY}

\author{\footnotesize ERNESTO A. MATUTE}

\address{Departamento de F\'{\i}sica, Universidad de Santiago de
Chile,\\ Usach, Casilla 307 -- Correo 2, Santiago, Chile\\
ernesto.matute@usach.cl}

\maketitle

\pub{}{}

\begin{abstract}
Presymmetry, the hidden symmetry underlying the charge and
generational patterns of quarks and leptons, is utilized for
repairing the left--right asymmetry of the standard model with
Dirac neutrinos. It is shown that the restoration of parity is
consequent with an indispensable left--right symmetric residual
presymmetry. Thus, presymmetry substantiates left--right symmetry
and the experimental search for the latter is the test of the
former, with the nature of neutrinos as a crucial feature that can
distinguish the left--right symmetry alone and its combination
with presymmetry. This phenomenological implication is in
accordance with the fact that Majorana neutrinos are usually
demanded in the first case, but forbidden in the second.

\keywords{Parity symmetry; charge symmetry; flavor symmetry;
presymmetry.}
\end{abstract}

\ccode{PACS Nos.: 11.30.Er, 11.30.Hv, 11.30.Ly, 12.60.Cn}

\section{Introduction}

The phenomenological success of the standard model (SM) of
particle physics based on the gauge group SU(3)$_c$ $\times$
SU(2)$_L$ $\times$ U(1)$_Y$ is moderated by a number of problems.
There is no logical sense for the complete asymmetry between left
and right in the weak sector and no explanation for the charge
symmetry between quarks and leptons. It also offers no reason for
the existence of fermion family copies and no prediction for their
numbers. Faced with these troubles, many theoretical ideas have
been advanced beyond the SM.

The minimal extension of the SM which repairs its left--right (LR)
asymmetry \linebreak is in the LR symmetric models (LRSM) with
gauge group $\mbox{SU(3)}_c \times \mbox{SU(2)}_L \times
\mbox{SU(2)}_R \times \mbox{U(1)}_{B-L}$, where an interchange
symmetry between the left and right sectors is
assumed.\cite{Mohapatra} The other popular choice to rectify the
LR asymmetry of the SM is held by the mirror matter models based
on the gauge group $G_L \times G_R$, where $G_{L,R} =
\mbox{SU(3)}_c \times \mbox{SU(2)}_{L,R} \times \mbox{U(1)}_Y$,
with equal gauge coupling constants for the symmetric
sectors.\cite{Foot1}$^{\mbox{--}}$\cite{Foot3} However, none of
these extended chiral models with LR symmetry finds the solutions
to the quark--lepton U(1)-charge symmetry and family problems.

These issues indeed have been addressed within the SM itself via
presymmetry, an electroweak symmetry between quarks and leptons
with Dirac neutrinos hidden by the nontrivial topology of weak
gauge fields.\cite{EAM1,EAM2} Our purpose in this letter is to
consider the possible LR symmetric extensions of presymmetry
beyond the SM in order to have a testable residual presymmetry
with LR symmetry as in the LRSM and mirror matter models, and
therefore give unified answers to the important fundamental
questions on charge symmetries, triplication of families and LR
asymmetry left open by the SM. In Sec.~2, we refer to presymmetry
within the context of the SM, emphasizing relevant points to this
work. In Sec.~3, we deal with the LR symmetric extension of
presymmetry leading to the LR symmetry embedded in the LRSM,
distinguishing phenomenologically the conventional models and
those supplemented with presymmetry. The alternative residual
presymmetry connected with mirror matter models was put forth in
Ref.~\refcite{EAM3} and the essential results are confronted here
with those related to the LRSM. The conclusions are presented in
Sec.~4.

\section{On Presymmetry in the SM}
\label{SM}

The basis of presymmetry is an electroweak quark--lepton symmetry
within the SM. For a weak hypercharge defined in terms of the
electric charge and the third component of weak isospin as
$Q=T_{3L}+Y/2$, such a symmetry can be read in the following
chiral relations\cite{EAM1,EAM2}:
\begin{equation}
Y(q_{L,R}) = \displaystyle Y(\ell_{L,R}) + \Delta Y (\ell_{L,R}) ,
\qquad Y(\ell_{L,R}) = \displaystyle Y(q_{L,R}) + \Delta Y
(q_{L,R}) ,
\label{chargesym}
\end{equation}
where $\Delta Y$ involves the baryon and lepton numbers according
to
\begin{equation}
\Delta Y = - \frac{4}{3} \, (3B-L)
\label{Delta1}
\end{equation}
and $q_{L,R}$ and $\ell_{L,R}$ refer to the quark and lepton weak
partners in $L$-doublets and $R$-singlets of SU(2)$_L$ within each
of the three families of the SM, right-handed neutrinos of $Y=0$
included; parity symmetry is broken in SU(2)$_L$ and U(1)$_Y$. The
hypercharge normalization can change the value of the global
fractional part $\Delta Y$, with the 3 attributable to the number
of quark colors, but not the underlying charge symmetry.

Presymmetry emerges from the correspondence of quark and lepton
charges if the global piece is kept away, easily perceived in
Eq.~(\ref{chargesym}). This quark--lepton symmetric pattern makes
sense only for Dirac neutrinos.

To understand the charge symmetry and the charge dequantization
hidden in Eq.~(\ref{chargesym}), the prequark (prelepton) states
denoted by $\hat{q}$ ($\hat{\ell}$) are introduced. They are
defined by the quantum numbers of quarks (leptons), except charge
values. Hypercharges of prequarks (preleptons) are the same as
their lepton (quark) weak partners. From Eq.~(\ref{chargesym}) one
is led to
\begin{equation}
Y(q_{L,R}) = Y(\hat{q}_{L,R}) + \Delta Y (\hat{q}_{L,R}) , \qquad
Y(\ell_{L,R}) = Y(\hat{\ell}_{L,R}) + \Delta Y (\hat{\ell}_{L,R})
,
\label{hathyper}
\end{equation}
where now
\begin{equation}
\Delta Y = - \frac{4}{3} \, (B-3L) .
\label{Delta2}
\end{equation}
Here the combination $B-3L$ is instead of $3B-L$ because prequarks
(preleptons) are entities that take the lepton (quark) hypercharge
values. This implies $B(\hat{q}_{L,R})=-1$ and
$L(\hat{\ell}_{L,R})=-1/3$, with the 3 attributable to the number
of families.\cite{EAM2}

The charge symmetry in Eq.~(\ref{chargesym}) and the charge
dequantization in Eq.~(\ref{hathyper}) are kept up with $3B-L$ and
$B-3L$ as ungauged global symmetries, quarks and leptons as the
ultimate constituents of ordinary matter, and prequarks and
preleptons as their basic bare states.

The hidden quark--lepton charge symmetry is implemented under the
premise that the global piece of hypercharge has a weak
topological character. Since any weak topological feature cannot
have observable effects at the zero-temperature scale because of
the smallness of the weak coupling, the charge structure reflected
in Eq.~(\ref{hathyper}) does not apply to quarks, but to new
states referred to as topological quarks. Nonetheless the
assignments of topological quarks to the gauge groups of the SM
are the same of quarks. The electroweak presymmetry is indeed
between topological quarks and preleptons having nontrivial charge
structure, and between prequarks and leptons with no charge
structure.

The interactions of prequarks (topological quarks) and leptons
(preleptons) with the gauge and Higgs fields are assumed to be
described by the same Lagrangian of the SM with quarks and leptons
except hypercharge couplings and inclusion of Dirac neutrinos.

The nonstandard fermionic hypercharges generate the
$\mbox{U(1)}[\mbox{SU(2)}]^2$ and $[\mbox{U(1)}]^3$ gauge
anomalies in the couplings by fermion triangle loops of three
currents associated with the chiral U(1) and SU(2) gauge
symmetries. Their cancellations require a counterterm which
includes topological currents or Chern--Simons classes related to
the U(1) and SU(2) gauge groups, leading to the appearance of
nontrivial topological winding numbers in the case of pure gauge
fields SU(2). Vacuum states labelled by different topological
numbers are then tunneled by SU(2) instantons carrying topological
charges, which are responsible for the transitions and charge
shifts from the nonstandard to standard fermions. Explicitly, in
the presymmetric scenario of prequarks and leptons, for instance,
each prequark hypercharge is equally modified by an amount set
down as\cite{EAM1,EAM2}:
\begin{equation}
\Delta Y (\hat{q}_{L,R}) = - \frac{n}{3} \; (B-3L)(\hat{q}_{L,R})
, \label{norma}
\end{equation}
where $n$ is the topological charge of a weak SU(2) instanton.

The anomaly cancellation and removal of the related counterterm,
needed for gauge invariance and renormalizability of the theory,
fix the index $n=4$ in Eq.~(\ref{norma}). This value matches
Eq.~(\ref{norma}) to (\ref{Delta2}) and is independent of the
normalization used for hypercharge.\cite{EAM3}

\section{LR Symmetry as a Residual Presymmetry}

The transitions from prequarks to topological quarks and from the
latter to quarks through weak SU(2) instantons do not take place
in the real world because, as argued in Ref.~\refcite{EAM3},
prequarks and topological quarks, in addition to preleptons, are
not real dynamical entities. They are bare prestates of quarks and
leptons seen as mathematical entities out of which the actual
particle states are built up. In a sense, such transitions are
truncated by the extreme smallness of the instanton transition
probability.

In spite of the fact that presymmetry is a hidden electroweak
symmetry within the system of quarks and leptons, the scheme
defines a theoretical framework with several physical
implications\cite{EAM2}: it explains the fractional charge of
quarks and the quark--lepton charge relations, it correlates the
number of fermion generations with the number of quark colors, and
it predicts the Dirac character of neutrinos. Also, it accounts
for the topological charge conservation in quantum flavor
dynamics, and it explains charge quantization and the no
observation of fractionally charged hadrons.

Yet, there is nothing physically new at the level of quarks and
leptons, so that the quark--lepton hidden presymmetry cannot be
tested. In this case, it is impossible to either verify or falsify
the underlying symmetric scheme, which is really difficult to
accept. These are strong motivations for a verifiable residual
presymmetry via partial or complete duplication of standard
particles, in a LR symmetric manner to have unified answers to
questions left open by the SM.

The basic premise is that presymmetry between quarks and leptons
is LR symmetric, doubling the weak gauge group of the SM in the
case of a minimal extension, just as in the LRSM. They are
arranged in left- and right-handed doublets of SU(2)$_L$ and
SU(2)$_R$, respectively\cite{Mohapatra}: $q_{L,R}=(u,d)_{L,R}$ and
$\ell_{L,R}=(\nu,e)_{L,R}$, within each of their three families.
The formula for the electric charge is $Q=T_{3L}+T_{3R}+(B-L)/2$
and the quark--lepton charge symmetry is as in
Eq.~(\ref{chargesym}) with $B-L$ instead of $Y$; it is the gauged
combination of $3B-L$ and $B-3L$ global symmetries.

The scheme to understand this charge symmetry and the charge
dequantization like in Eq.~(\ref{hathyper}) is similar to that in
the SM, described in Sec.~\ref{SM}. In the scenario of symmetric
prequarks and leptons, there are two $\mbox{U(1)}[\mbox{SU(2)}]^2$
gauge anomalies and the charge shifts of Eq.~(\ref{norma}) from
prequarks to topological quarks are now
\begin{equation}
\displaystyle \Delta (B-L) (\hat{q}_{L,R}) = - \frac{n}{3} \;
(B-3L)(\hat{q}_{L,R}) ,
\label{newnorma}
\end{equation}
where $n=n_L-n_R$, with $n_{L,R}$ being the topological charge of
the SU(2)$_{L,R}$ instanton; the change of sign is due to the
axial character of the anomaly. The requirements of anomaly
cancellation, removal of the corresponding counterterm and LR
symmetry lead to $n_L=-n_R=2$, making Eq.~(\ref{newnorma}) equal
to (\ref{Delta2}) with $B-L$ in place of $Y$. For simplicity, we
indicate this hidden charge symmetry relating quark and lepton
doublets as
\begin{equation}
q_L \leftrightarrow \ell_L , \qquad q_R \leftrightarrow \ell_R .
\end{equation}

Similarly, due to the vectorial character of $B$ and $L$, there is
the hidden presymmetry given by
\begin{equation}
q_L \leftrightarrow \ell_R , \qquad q_R \leftrightarrow \ell_L ,
\end{equation}
with symmetry between the SU(2)$_L$ and SU(2)$_R$ gauge and Higgs
fields of equal couplings, and the underlying topological-charge
symmetry $n_L=-n_R=2$.

These hidden $Z_2$ symmetries introduce LR symmetry in
presymmetry, leading to the usual LR symmetry
\begin{equation}
q_L \leftrightarrow q_R , \qquad \ell_L \leftrightarrow \ell_R ,
\end{equation}
which remains exact after the underlying charge normalization with
the topological-charge correspondence $n_L \leftrightarrow -n_R$.
Thus, parity is restored and the LR symmetry itself embedded in
the LRSM becomes the required testable residual presymmetry.

An alternative residual presymmetry manifested as a LR symmetry is
provided by mirror matter models, put forth in Ref.~\refcite{EAM3}
and included here for completeness. Other motivations to consider
this possibility are to extend presymmetry from matter to forces
and from the electroweak to the strong sector, doubling the
particle spectrum completely.

On the one hand, it is the electroweak presymmetry relating chiral
quarks and leptons within each of their three families, as
explained in Sec.~\ref{SM}, and their respective mirror partners
denoted by tildes:
\begin{eqnarray}
\begin{array}{l}
q_L \leftrightarrow \ell_L , \qquad u_R \leftrightarrow \nu_R ,
\qquad d_R \leftrightarrow e_R , \\ [12pt] \tilde{q}_R
\leftrightarrow \tilde{\ell}_R , \qquad \tilde{u}_L
\leftrightarrow \tilde{\nu}_L , \qquad \tilde{d}_L \leftrightarrow
\tilde{e}_L ,
\end{array}
\label{Presymm}
\end{eqnarray}
with charge relations as in
Eqs.~(\ref{chargesym})--(\ref{Delta2}), and similarly for the set
of mirror partners. The underlying charge symmetry is hidden by
the shifts produced by the topological charges of weak instantons.
Regarding these, we note that the analogous of Eqs.~(\ref{norma})
and (\ref{newnorma}) takes the form
\begin{eqnarray}
\begin{array}{l}
\displaystyle \Delta Y (\hat{q}_{L,R}) = - \frac{n_L}{3} \;
(B-3L)(\hat{q}_{L,R}) , \\ [12pt] \displaystyle \Delta Y
(\hat{\tilde{q}}_{L,R}) = + \frac{n_R}{3} \;
(B-3L)(\hat{\tilde{q}}_{L,R}) ,
\end{array}
\label{mirror}
\end{eqnarray}
where $q$ and $\tilde{q}$ refer to quarks and partners in
Eq.~(\ref{Presymm}). The values of the topological charges
demanded by the cancellation of the $\mbox{U(1)}[\mbox{SU(2)}]^2$
and $[\mbox{U(1)}]^3$ anomalies, and the related counterterms, are
$n_L=4$ for prequarks and $n_R=-4$ for their partners, with the
correspondence $n_L \leftrightarrow -n_R$. These results match
Eq.~(\ref{mirror}) to (\ref{Delta2}) for prequarks and mirror
partners. Here gauge and Higgs fields are not interchanged by the
presymmetric invariant transformations in the electroweak sectors.

On the other hand, there is a similar hidden charge symmetry
between quarks and the partners of leptons, and between their
copies, respectively:
\begin{eqnarray}
\begin{array}{l}
q_L \leftrightarrow \tilde{\ell}_R , \qquad u_R \leftrightarrow
\tilde{\nu}_L , \qquad d_R \leftrightarrow \tilde{e}_L , \\ [12pt]
\tilde{q}_R \leftrightarrow \ell_L , \qquad \tilde{u}_L
\leftrightarrow \nu_R , \qquad \tilde{d}_L \leftrightarrow e_R .
\end{array}
\label{partnerPresymm}
\end{eqnarray}
The underlying charge relations are as in Eq.~(\ref{Presymm}), but
interchanging $\ell_L \leftrightarrow \tilde{\ell}_R$, \linebreak
$\nu_R \leftrightarrow \tilde{\nu}_L$, $e_R \leftrightarrow
\tilde{e}_L$. The electroweak symmetry which now interchanges the
gauge and Higgs bosons with their partners requires that the
corresponding coupling constants be equal.

The $Z_2$ symmetries of Eqs.~(\ref{Presymm}) and
(\ref{partnerPresymm}) put LR symmetry into presymmetry, implying
the following one:
\begin{eqnarray}
\begin{array}{l}
q_L \leftrightarrow \tilde{q}_R , \qquad u_R \leftrightarrow
\tilde{u}_L , \qquad d_R \leftrightarrow \tilde{d}_L , \\ [12pt]
\ell_L \leftrightarrow \tilde{\ell}_R , \qquad \nu_R
\leftrightarrow \tilde{\nu}_L , \qquad e_R \leftrightarrow
\tilde{e}_L ,
\end{array}
\label{resPresymm}
\end{eqnarray}
with the same coupling constants for electroweak gauge and Higgs
bosons and their partners. This discrete symmetry, but not the two
others, remains exact after the underlying charge normalization on
fermions. Moreover, it extends to strong interactions for equal
gauge couplings of the two color groups. A residual $Z_2$ symmetry
then appears, which includes the strong sector, relates every
particle with its partner and constrains the corresponding
coupling constants to be equal, restoring parity symmetry, just as
mirror symmetry in mirror matter models. Consequently, mirror
symmetry can be regarded as the verifiable residual presymmetry
that is requested.

Experimentally testable predictions are obtained from the LRSM or
mirror matter models with Dirac neutrinos. These are the
phenomenological implications of combining LR symmetry with
presymmetry. In the presymmetry model, neutrinos are of Dirac
type. But in the LRSM and mirror matter models, they can be both
Dirac and Majorana types in principle, with Majorana neutrinos in
connection with seesaw mechanisms being the more popular choices.
Thus, the nature of neutrinos is one of the phenomenological
features distinguishing the conventional LRSM and mirror matter
models in which Majorana neutrinos are allowed and those
supplemented with presymmetry in which these are forbidden.

\section{Conclusions}

The charge symmetries between quarks and leptons are explained by
the hidden presymmetry underlying the displayed patterns. Yet, it
is expanded beyond the SM in order to have an experimentally
testable residual presymmetry via partial or complete duplication
of standard particles\,---\,in a LR symmetric way for the purpose
of restoring parity and therefore solving puzzles of the SM
coherently. The doubling is implied by presymmetry apart from the
ultraviolet completion of the theory, which substantiates LR
symmetry. This includes the nightmare scenario with no Higgs
boson.

The relation between presymmetry and LR symmetry is shown to be so
robust that the experimental search for possible parity
restoration through LRSM\cite{Mohapatra} or mirror matter
models\cite{Foot1}$^{\mbox{--}}$\cite{Foot3} with Dirac neutrinos
is a test of presymmetry. This type of neutrinos discriminates,
phenomenologically, between the LRSM and mirror matter models
supplemented with presymmetry and those without it in which
Majorana neutrinos are not forbidden.

An experimental evidence for LR symmetry with Dirac neutrinos also
mean a corroboration of the answer provided by presymmetry to one
of the most intricate problems of elementary particle physics: the
triplication of families. This is the result of the organizing
presymmetric principles for quarks and leptons which correlates
the number of families with the number of quark color
charges.\cite{EAM2}

\section*{Acknowledgments}

This work was supported by the Departamento de Investigaciones
Cient\'{\i}ficas y Tecnol\'ogicas, Universidad de Santiago de
Chile, Usach.

\end{document}